\documentclass[3p,times]{elsarticle}

\usepackage{ecrc}

\usepackage{epstopdf}

\volume{00}

\firstpage{1}

\journalname{Nuclear Physics A}

\runauth{Frash\"er Loshaj et al.}

\jid{nupha}

\jnltitlelogo{Nuclear Physics A}

\usepackage{graphicx}
\usepackage{amsmath,amssymb}
\begin{document}

\begin{frontmatter}



\title{Soft photon production \\ from real-time dynamics of jet fragmentation}

%% Single author (and collaboration) - please insert
%\author{Frash\"er Loshaj}
%\fntext[col1] {A list of members of the XYZ Collaboration and acknowledgements can be found at the end of this issue.}
%\address{Department of Physics and Astronomy, Stony Brook University, Stony Brook, New York 11794-3800, USA}

%% For multiple authors, replace the above by:

\author[label1]{Frash\"er Loshaj}
\author[label1,label2]{Dmitri E. Kharzeev}

\address[label1]{Department of Physics and Astronomy, Stony Brook University, Stony Brook, New York 11794-3800, USA}
\address[label2]{Department of Physics, Brookhaven National Laboratory, Upton, New York 11973-5000, USA}
%\address[label2]{Department of Physics, Brookhaven National Laboratory, Upton, New York 11973-5000, USA}

\begin{abstract}
Soft photons produced in heavy ion collisions are an important tool for probing the properties of the quark-gluon plasma. It is therefore crucial to understand the background - soft photons produced in elementary collisions. Low theorem states that soft photon production in hadron collisions is dominated by Bremsstrahlung off charged initial and final state hadrons. Surprisingly, almost every experiment observed an enhancement (by a factor of $2\div 5$) above Low theorem's prediction. This is the longstanding puzzle of ``anomalous soft photon production.'' The phenomenon is not observed in processes with leptonic final states, which suggests that the mechanism is due to nonperturbative QCD evolution. We study this phenomenon using an exactly soluble, massless, Abelian model in $1+1$ dimensions which shares with QCD many important properties: confinement, chiral symmetry breaking, axial anomaly and $\theta$-vacuum. We then apply this model to the soft photon production in the fragmentation of jets produced in $Z^0$ decays and find a qualitative agreement with the data. 
\end{abstract}


\end{frontmatter}

% \linenumbers


\section{Introduction}
\label{intro}
From the Low theorem \cite{Low:1958sn} it follows that soft photon production in hadron collisions is dominated by Bremsstrahlung off incoming and outgoing charged hadrons. Surprisingly, almost every experiment observed an enhancement by a factor of $2\div 5$ above Low theorem's predictions (see for example \cite{Chliapnikov:1984ed,Botterweck:1991wf,Banerjee:1992ut,Belogianni:1997rh,Belogianni:2002ib,Belogianni:2002ic}). This is the long standing puzzle of ``anomalous photon production.'' Many theoretical models have been proposed to explain this phenomenon (for a review of earlier work, see e.g. \cite{Balek:1989rx}; see also \cite{Shuryak:1989vn,Lichard:1990ye,Botz:1994bg,Wong:2010gf,Hatta:2010kt}), however, none of them explains all the features of it. 
Recently, photon yield was measured in hadronic decays of the $Z^0$ boson by the DELPHI Collaboration \cite{Abdallah:2005wn,Abdallah:2010tk}. It is observed that the photon spectrum is similar in shape to what is expected from Bremsstrahlung, but with a magnitude about {\it four} times higher. Such a discrepancy is not observed \cite{Abdallah:2007aa} for photons produced in $Z^0$ decays to $\mu^+\mu^-$. This suggests that the enhancement of soft photons comes from the intermediate stage of hadron formation \cite{Kharzeev:2013wra}. 
%
%Understanding the mechanism of anomalous soft photon production is important for using soft photons to probe the properties of quark-gluon plasma, as we need to %understand the ``background,'' the mechanism of photon production in elementary collisions.

As the leading quark of the jet moves in the QCD vacuum, it pulls from the Dirac sea pairs of quark-antiquark, which later hadronize. Quarks are electrically charged and generate an electromagnetic current during this process that can be considered to be a new source of photon production. This involves nonperturbative dynamics that is not yet understood in QCD from first principles, therefore we rely on a model. Based on the picture of confinement due to condensation of magnetic monopoles \cite{'tHooft:1977hy,Mandelstam:1974pi} (dual Meissner effect), we assume that the dynamics along the jet axis is Abelian. We also assume dimensional reduction due to the large energies of jets considered. Based on these assumptions, similarly as in \cite{Loshaj:2011jx,Kharzeev:2012re} we assume that the longitudinal dynamics of the jet is described by massless ${\rm QED}_2$, known as the Schwinger model \cite{Schwinger:1962tp,Lowenstein:1971fc,Coleman:1975pw}. 

\section{Dynamics of jet fragmentation}
\label{dynjet}
\begin{figure}[htbp]
\centering
\includegraphics[width=.43\linewidth]{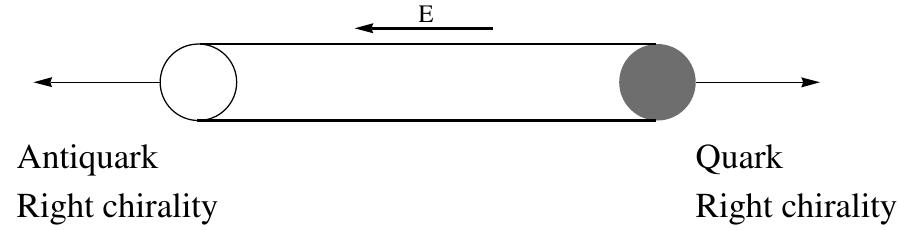} \hspace{15 mm}
\includegraphics[width=0.36\linewidth]{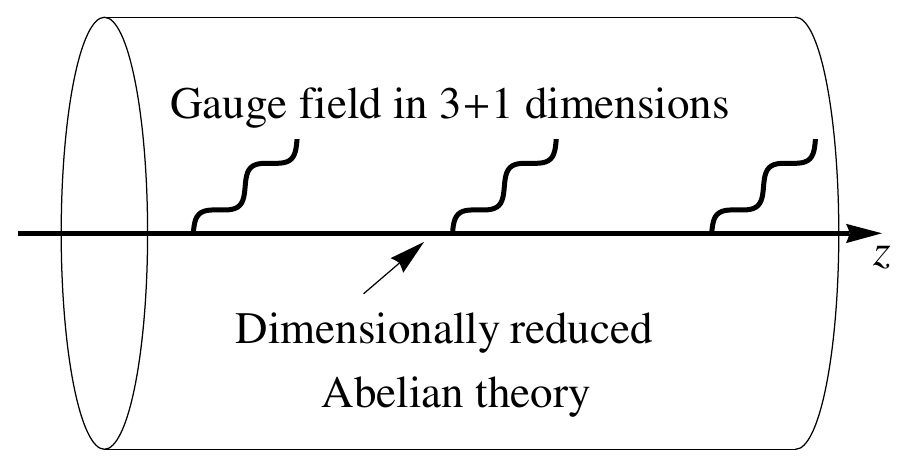}
\caption{Quark - antiquark moving back-to-back, left, and electrically charged $1+1$ current, coupled to a $3+1$ dimensional electromagnetic field, right.}
\label{fig:btbchrg}
\end{figure}

The dynamics of ${\rm QED}_2$, also known as the Schwinger model \cite{Schwinger:1962tp}, is described by the Lagrangian:
\begin{equation}
\mathcal{L}=-\frac{1}{4}G_{\mu\nu} G^{\mu\nu}+\bar{\psi}i\gamma^\mu \partial_\mu \psi-g \bar{\psi}\gamma^\mu \psi B_\mu - g j_{ext}^\mu B_\mu
\label{eq:qed2}
\end{equation}
where $\mu=0,1$, $B_\mu$ is the gauge field, $G_{\mu\nu}=\partial_\mu B_\nu-\partial_\nu B_\mu$ is the field strength and $j_{ext}^\mu$ is an external current. The di-jet is introduced in the theory via the charge density \cite{Loshaj:2011jx,Kharzeev:2012re} (see Figure \ref{fig:btbchrg}, left):

\begin{equation}
j_{ext}^0(x)=\delta(z-vt)\theta(z)-\delta(z+vt)\theta(-z),
\label{eq:ec}
\end{equation}
where $v=\frac{p_{jet}}{\sqrt{p_{jet}^2+Q_0^2}}$, $p_{jet}$ is the jet momentum and $Q_0 \sim 2$ GeV is an infrared scale, roughly corresponding to the scale at which the pQCD DGLAP cascade stops, and the effects of confinement described by our effective theory begin to operate. We have not considered the emissions of the additional partons in the DGLAP cascade.

It is well known that the Schwinger model can be solved by bosonisation \cite{Coleman:1974bu,Mandelstam:1975hb}. 
For the vector current, the bosonization relation is
\begin{equation}
j^\mu(x) = \bar{\psi}(x)\gamma^\mu \psi(x)=-\frac{1}{\sqrt{\pi}}\epsilon^{\mu\nu}\partial_\nu\phi(x),
\label{eq:bc}
\end{equation}
where $\phi$ is a real scalar field. It can be shown that the equation of motion for the scalar field is \cite{:1974cks,Loshaj:2011jx,Kharzeev:2012re}
\begin{equation}
(\Box+m^2)\phi(x)=-m^2\phi_{ext}(x)
\label{eq:eom}
\end{equation}
where $\Box \equiv \partial_t^2 - \partial_z^2$ and $m^2=g^2/\pi$ and $j_{ext}^\mu(x)=-\frac{1}{\sqrt{\pi}}\epsilon^{\mu\nu}\partial_\nu\phi_{ext}(x)$. We use the value $m=0.6 \ \mathrm{GeV}$ \cite{Loshaj:2011jx,Kharzeev:2012re}.

The coupling to a classical source results in particle creation. The produced particles can be interpreted as neutral mesons produced in the fragmentation of the string stretched between the original quark and antiquark. The induced electric field is given by $F_{01}=-\frac{g}{\sqrt{\pi}}\phi$. In the case of quark-antiquark moving with the speed of light, $v=1$, one can calculate the induced axial charge as a function of time \cite{Kharzeev:2013wra} considering the total, i.e. external and induced, field $F^{tot}_{01}$
\begin{equation}
\Delta Q_5=\frac{g}{\pi}\int{d^2 x F_{01}^{tot}}=2\left[\cos(mt)-1\right].
\label{eq:}
\end{equation} 
This leads, via anomaly, to the oscillation of vector current and in turn to photon production.
 
\section{Soft photon production}
\label{softph}

The induced field $\phi$ due to the external source of quark-antiquark can be computed from \eqref{eq:eom} and using bosonization relation \eqref{eq:bc}, one can compute the induced electromagnetic current, by multiplying it by the fractional charge of the quark of flavor $f$, $Q_f e$. The electromagnetic field is not confined along the jet axis, and can propagate in $(3+1)$  dimensions (see Figure \ref{fig:btbchrg}, right). Denoting by $j^\mu_{tot}$ the electromagnetic current generated from the original quark-antiquark and the induced field $\phi$, the photon Bremsstrahlung spectrum can now be evaluated using the standard formula
\begin{equation}
\frac{dN_\gamma}{d^3 p}=\frac{1}{(2\pi)^3}\frac{1}{2 p^0}|\tilde{j}_{tot}^\mu(p) \tilde{j}^*_{tot,\mu}(p)| ,
\label{eq:tpp}
\end{equation}
where $p^0=|\bold{p}|$ and $\tilde{j}_{tot}$ is the Fourier transform of the current. $p^\mu$ is the $3+1$ momentum, so we make the identification $p_\mu p^\mu=p_0^2-p^2_z= p_\perp^2$. We have to also introduce the probability for $Z_0$ to decay to a certain flavor of quark. The final result can be written as \cite{Kharzeev:2013wra}:
\begin{eqnarray}
\frac{dN_\gamma}{d^3 p}=\left(\frac{\Gamma_{uu}+\Gamma_{cc}}{\Gamma_{\mathrm{hadron}}}\left(\frac{2}{3}\right)^2+\frac{\Gamma_{dd}+\Gamma_{ss}+\Gamma_{bb}}{\Gamma_{\mathrm{hadron}}}\left(\frac{1}{3}\right)^2\right)\frac{1}{(2\pi)^3}\frac{1}{2 p^0}e^2 \frac{4 v^2}{(p_0^2-v^2 p_z^2)^2}p_\perp^2\left(1+\frac{m^2}{p_\perp^2-m^2}\right)^2 %\nonumber \\
%&=&\left(B_{2/3}\left(\frac{2}{3}\right)^2+B_{1/3}\left(\frac{1}{3}\right)^2\right)\frac{1}{(2\pi)^3}\frac{1}{2 p^0}e^2 \frac{4 v^2}{(p_0^2-v^2 %p_z^2)^2}p_\perp^2\left(1+\frac{m^2}{p_\perp^2-m^2}\right)^2
\label{eq:tpp1}
\end{eqnarray}
where $\Gamma_{ff}$ is the decay width of $Z_0$ to quark-antiquark of flavor $f$ and $\Gamma_{\mathrm{hadron}}$ is the total decay width of $Z_0$ to hadrons; the Particle Data Group \cite{Beringer:1900zz} gives the values $\frac{\Gamma_{uu}+\Gamma_{cc}}{\Gamma_{\mathrm{hadron}}}=0.331$ and $\frac{\Gamma_{dd}+\Gamma_{ss}+\Gamma_{bb}}{\Gamma_{\mathrm{hadron}}}=0.669$. 

It is important to note that in the soft photon limit of $p_\perp \to 0$, the two terms in the parenthesis cancel each other. This means that the soft limit is in accord with the Low theorem, because we only have neutral mesons.
 
%\begin{figure}
%\centering
%\includegraphics[width=0.4\linewidth]{./seeqtogfield_new.pdf} 
%\caption{Electrically charged $1+1$ current, resembling a quantum wire, coupled to a $3+1$ dimensional electromagnetic field.}
%\label{fig:seeqtogfield_new}
%\end{figure} 

The model gives a sharp enhancement for $p_\perp=m$. To get a more realistic result, we have to consider a distribution in $m$, as the scalar meson of our effective theory represents the entire hadron spectrum, and finite width of the mesons. 

For a fermion-antifermion pair separated by the distance $r$, the potential at short distances $r \ll m^{-1}$ is linear, 
\begin{equation}
V(r) \simeq \frac{\pi}{2}m^2 r, \ \ r \ll m^{-1} ,
\label{eq:pot}
\end{equation}
with the string tension $\kappa^2=\frac{\pi}{2}m^2$.  Let us assume that the string tension fluctuation is Gaussian \cite{Bialas:1999zg}
\begin{equation}
P(\kappa^2)=\sqrt{\frac{2}{\pi <\kappa^2>}}e^{-\frac{\kappa^2}{2 <\kappa^2>}}.
\label{eq:}
\end{equation} 
We use $<\kappa^2> = 0.9$ GeV/fm suggested by the lattice studies and Regge phenomenology. We also assume a finite width for the scalar propagator, and the term in parentheses in \eqref{eq:tpp1} becomes
\begin{equation}
\left|1+\frac{m^2}{p_\perp^2-m^2+i\gamma^2}\right|^2=\frac{p_\perp^4+\gamma^4}{(p_\perp^2-m^2)^2+\gamma^4}\underset{\gamma<<m}{\rightarrow} \left(1+\frac{p_\perp^4}{\gamma^4}\right)\left(\gamma^2\frac{ \pi}{2m}\right)\delta(p_\perp-m),
\label{eq:df}
\end{equation}
where $\gamma=\sqrt{m \Gamma}$ is an effective width. Using the PDG \cite{Beringer:1900zz}  values for the masses and widths of the neutral isoscalar resonances, we find the values of $\gamma$ in the range of $\gamma \simeq 8\cdot 10^{-4}$ GeV for the $\eta$ meson, and $\gamma \simeq 8\cdot 10^{-2}$ GeV for the $\omega$ meson.

\begin{figure}
\centering
\includegraphics[width=0.58\linewidth]{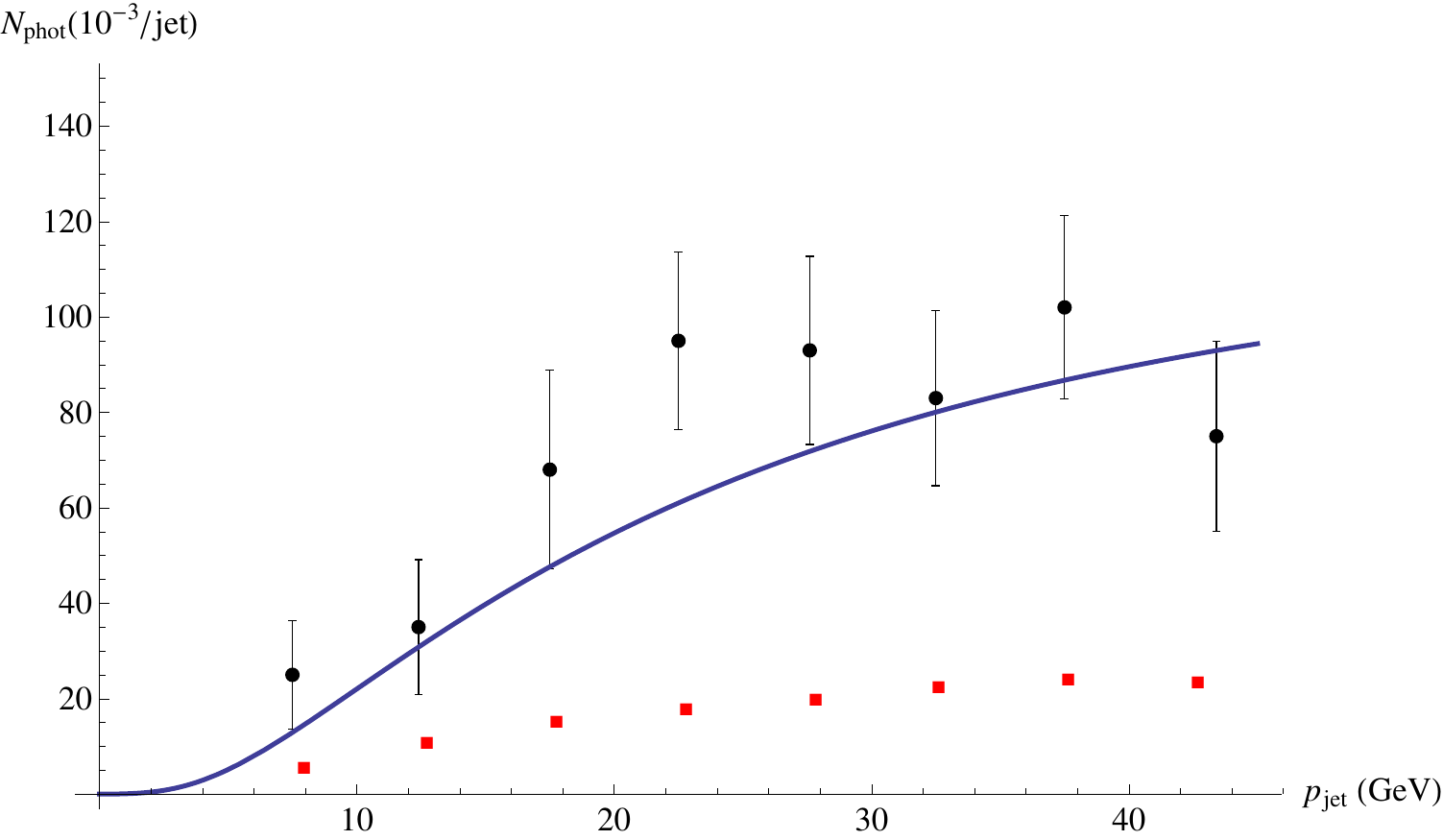} 
\caption{The soft photon yield as a function of the jet momentum. The circles represent the data and squares show calculations from Low theorem \cite{Abdallah:2010tk}. The solid line is our result.}
\label{fig:ngvspjet}
\end{figure}

The DELPHI Collaboration \cite{Abdallah:2010tk} measured the photons with transverse momenta $p_\perp<80$ MeV and total energies within $0.2<E_\gamma<1$ GeV. 
We compute
\begin{equation}
N_\gamma =\int{dm \ \sqrt{\frac{\pi}{2}} \ P\left(\frac{\pi}{2}m^2\right) \ \left(\int{d^3 p \frac{dN_\gamma}{d^3p}}\right) }  
\label{eq:total}
\end{equation} 
by integrating over the same range of transverse momentum and energy. We plot the result in Figure \ref{fig:ngvspjet} using $\gamma \simeq 0.003$ GeV. One can see that our mechanism describes the observed enhancement reasonably well, with the fitted value of the parameter $\gamma$ within a reasonable range expected for neutral isoscalar resonances.  

\section{Conclusion}
\label{concl}

Using a $1+1$ dimensional exactly soluble model, we proposed a possible mechanism of the observed phenomenon of soft photon production enhancement above Low theorem's predictions. The model is simple, but the justification for using it comes from the fact that it shares with QCD many important features, such as confinement, chiral symmetry breaking, anomalies and $\theta$-vacuum. In this model, as the jet fragments, it creates coherent oscillations of axial and vector currents, where the latter can be viewed as a source of photon emission. By introducing an adjustable parameter, we were able to describe the DELPHI data on the soft photon production. 

It would be interesting to generalize our study to a $3+1$ dimensional model. In heavy ion collisions, soft photons are very important for probing the properties of the quark gluon plasma and it is therefore crucial to account for the background coming from hard partons.

\textbf{Acknowledgments}: This work was supported by the U.S. Department of Energy under Contracts No. DE-FG-88ER40388 and No. DE-AC02-98CH10886. 



%% References


%\bibliographystyle{elsarticle-num}



\end{document}